# The Landscape of Academic Literature in Quantum Technologies

**Zeki C. Seskir[1*], Arsev U. Aydinoglu[2]**

*1- Physics Department, Middle East Technical University, Ankara, Turkey*

*2- Center for Science and Technology Policies, Middle East Technical University, Ankara, Turkey*

** Email address: zseskir@metu.edu.tr*

**Abstract**

In this study, we investigated the academic literature on quantum technologies (QT) using bibliometric tools. We used a set of 49,823 articles obtained from the Web of Science (WoS) database using a search query constructed through expert opinion. Analysis of this revealed that QT is deeply rooted in physics, and the majority of the articles are published in physics journals. Keyword analysis revealed that the literature could be clustered into three distinct sets, which are (i) quantum communication/cryptography, (ii) quantum computation, and (iii) physical realizations of quantum systems. We performed a burst analysis that showed the emergence and fading away of certain key concepts in the literature. This is followed by co-citation analysis on the 'highly cited' articles provided by the WoS, using these we devised a set of core corpus of 34 publications. Comparing the most highly cited articles in this set with respect to the initial set we found that there is a clear difference in most cited subjects. Finally, we performed co-citation analyses on country and organization levels to find the central nodes in the literature. Overall, the analyses of the datasets allowed us to cluster the literature into three distinct sets, construct the core corpus of the academic literature in QT, and to identify the key players on country and organization levels, thus offering insight into the current state of the field. Search queries and access to figures are provided in the appendix.

**Article Highlights**

- Academic literature in QT can be clustered into three sets; (i) quantum communication/cryptography, (ii) quantum computation, and (iii) physical realizations of quantum systems
- Identification of core corpus of this literature is done using bibliometric analysis, there are differences in topics between what is cited in the highly cited works and general literature
- Burst detection shows how the QT literature is evolving, emergence and fading away of concepts in the literature
- The field has been becoming more collaborative for the last three decades
- Co-authorship analysis shows strong relations between European and American institutions while Chinese and Japanese institutions are more prone to internal collaboration

## Introduction

Quantum Technologies (QT) is a field gaining attention from funding agencies. In 2018, two new initiatives with a scale of 1 billion euros each were approved by their respective legislative bodies in the EU (Riedel et al. 2019) and the US (Raymer and Monroe 2019). When combined with the programs in Canada (Sussman et al. 2019), Japan (Yamamoto et al. 2019), Australia (Roberson and White 2019), the UK (Knight and Walmsley 2019), Russia (Fedorov et al. 2019), and China (Zhang et al. 2019), QT is being recognized as a field of importance by most of the developed world. Although the programs and initiatives aforementioned are recent, the idea of using quantum technologies for some of the novelties being presented today is almost 60 years old, such as using "...not just circuits, but some system involving the quantized energy levels, or the interactions of quantized spins" to study the behavior of atoms on small scales, in 1959 (Feynman 1959). These ideas are later formalized into the concept of quantum computers by Feynman in 1982, and the phrase "second quantum revolution" has been coined since 2003 (Dowling and Milburn).

The idea of a 'second' quantum revolution emerges from the fact that the use of quantum technologies that allowed the Information Age, transistors, and lasers, are reaching their physical limits. Moore's Law cannot continue indefinitely since miniaturization of transistors has a lower bound at one to five nm scales. Similarly, to increase our control over communication and sensing processes, there is a need to master the manipulation of single photons. These two are beyond the generally accepted realm of early modern physics and require the adoption of full-blown quantum theories of their respective areas.

Historically, quantum technologies became an active field emanating from certain developments such as formulation of BB84 (Bennett and Brassard 1984) quantum key distribution protocol and Shor's algorithm (Shor 1994) for prime factorization, which can break the commercially used RSA encryption. Lately, large IT companies such as IBM, Google, Microsoft, Honeywell, and Alibaba have been competing in their pursuit of developing the first quantum computer that can solve a problem more efficiently than its classical counterpart without a doubt and declare quantum supremacy (Harrow and Montanaro 2017). Recently researchers at Google have published an article claiming supremacy (Arute et al. 2019). Regardless, the interest in QT and academic literature has been growing rapidly since the early 2000s.

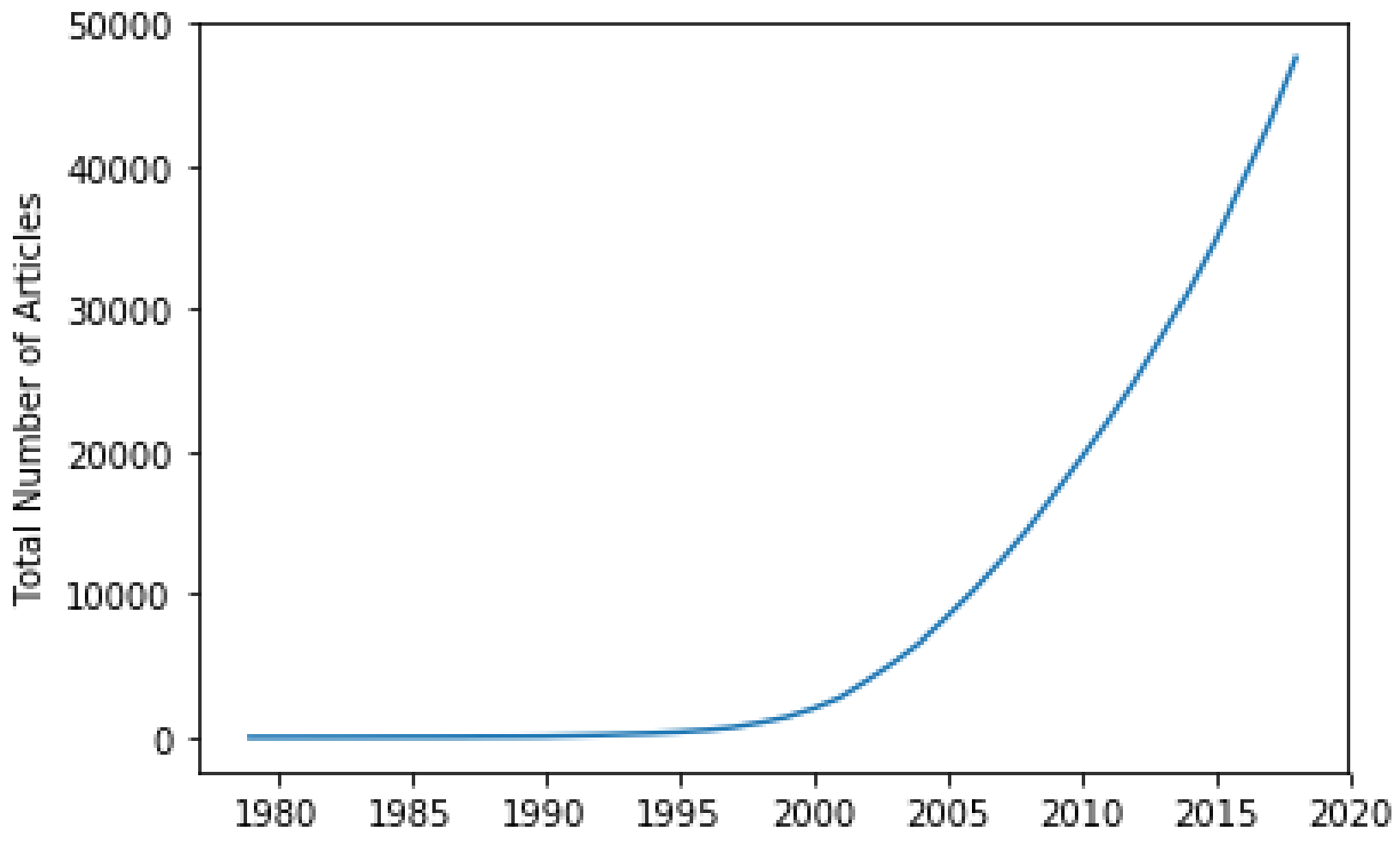

Figure 1: Number of total articles published in QT related research in WoS indexed databases (https://figshare.com/s/c114c3509e91d8c3484b)

The only studies on the academic literature in quantum technologies conducted using bibliometrics are of limited extent both in their search query and the time period and focuses only on specific sub-areas of quantum technologies (Olijnyk 2018; Dhawan, Gupta and Bhusan 2018). Considering the increasing amount of funding allocated recently to QT research, the elevated levels of activity in the field, and the rapid increase in academic work; it seems evident to us that an encompassing study employing bibliometric tools for a thematic analysis of the entire quantum technologies field is needed. With this study we aim to answer several research questions. (RQ1) As an academic field, which areas does QT research consist of? (RQ2) What are the emerging and disappearing research concepts in QT research? (RQ3) Has QT research become more collaborative over the years? (RQ4) What is the core corpus of the QT literature? (RQ5) Who are the important players in QT research?

**Methodology**

There have been several studies on different emerging fields using bibliometrics and visualization tools such as on quantum cosmology (Fay and Gautrias 2015), nanoscience (Porter and Youtie 2009; Luan and Porter 2017), astrobiology (Brazelton and Sullivan 2009; Taşkın and Aydinoglu 2015), origins of life (Aydinoglu and Taşkın 2018), neuroscience (Rosvall and Bergstrom 2010) and entrepreneurship (Chandra 2018). Our study follows a similar pattern to these studies.

As in most bibliometric studies, one of the key points is the determination of sample size and formulation of the query to target articles from their respective fields. In order to reach the right query to obtain articles falling within the academic literature of quantum technologies, we used a three-step procedure. Initially, we created a search query on the Web of Science database using personal expertise and certain common phrases gathered from policy documents such as quantum computing, quantum sensing, and so on. From this initial search, we collected 40,918 articles and proceeding papers in March 2019, and formulated keyword maps and burst analysis for emerging topics (query and main keyword map for the initial set can be found in Appendix). In the second step, we shared these with five area experts from the fields of quantum information theory, quantum optics, physics of information processing, and quantum thermodynamics. We obtained their feedback to enrich and widen our search query (see Appendix for the extended query). Finally, we generated keyword maps and burst analysis from this final set and compared it with our initial set.

With this new search query in June 2019, this time excluding the proceeding papers, we collected 49,823 records. The complete set was used for general analysis (WoS categories, research areas, journals, organizations, and countries). For further analysis, we curated the set through deleting academic work with zero TC (Web of Science Core Collection Times Cited Count) and NR (Cited Reference Count). After this process, there were 42,530 articles left.

Finally, to get an additional and focused view of the extensive academic literature, we obtained the Highly Cited articles presented by the Web of Science database for the specified search query. There were 808 articles from the period 1999-2018 with 234.59 average citations per item, and with a total number of 185,860 citations without self-citations. This set was used to form a core corpus for the literature on quantum technologies, which can be found in the Appendix section.

**Findings**

Using the initial set of 49,823 articles and Web of Science database internal tools, we have obtained the following lists; Web of Science Categories, Research Areas, and Source Titles (Journals). Through these, we have reached two descriptive qualities of the academic literature.

Around 85% of the articles were designated under Categories of physics and optics. Furthermore, out of the top 10 journals, seven of them were dedicated physics journals such as Physical Review A, and only two were journals focused on quantum information, the only journal with wide multidisciplinarity was Scientific Reports. In total, there were 495 journals with five or more articles from the total dataset. We have checked for the 80-20 rule (Lancaster and Lee 1985), also known as Pareto distribution, whether 80% of the articles are published at 20% of journals. This rule indicates whether a field is mature/stable or emergent, as emergent fields do not fit into the 80-20 rule. In the top 99 journals, there were 42,800 articles. Right now, 87% of articles are published in the top 20% of journals, which does not fit the expected distribution of a highly established field. Through these, we argue that the academic literature in quantum technologies is relatively established and mainly gathered around physics-related journals. Though in time, we expect that the field-specific journals are going to increase in number, and a similar distribution to a Pareto distribution will be observed.

**(RQ1) As an academic field, which areas does QT research consist of?**

We have continued our analysis with the remaining 42,530 articles in the curated set after this point. On this dataset, we have performed the co-occurrence of keyword analysis using VOSviewer, limiting ourselves to the top 60% of keywords occurring 200 times or more, in the title or abstract section of the article. The resulting map is presented in Figure 2.

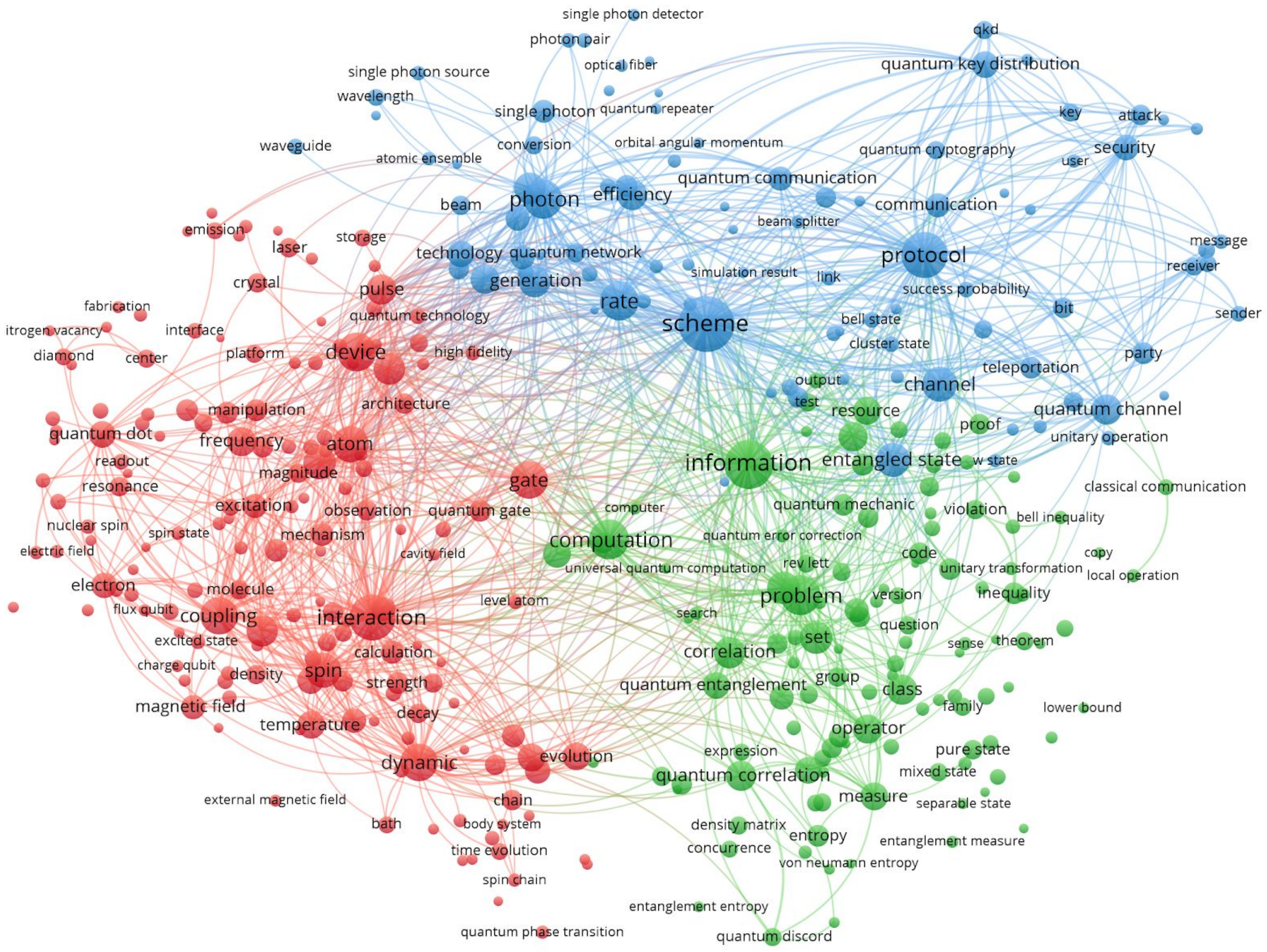

Figure 2: Clustering of terms occurring more than 200 times in the curated set (60% version)

(https://figshare.com/s/979baf88ff1ba55bc581)

There were three clusters. The red cluster mostly corresponds to the physical realizations while the blue cluster is quantum cryptography and communication, and the green cluster has quantum computation together with quantum information theory concepts. Keywords such as quantum discord, concurrence, relative entropy, and entanglement measure appear at the edges of this cluster. Indicating that, in the literature, articles on the more theoretical side of quantum information are more closely related to quantum computation rather than quantum cryptography, communication, or physical realization of these theoretical concepts.

A possible explanation of this consolidation of clusters was the exclusion of articles with TC and NR zero, which are the articles with no citations and references. A quick analysis over the entire set of 49,823 articles showed that the three clusters persist if the same procedure is followed with the reduced set. However, when the entire set (100%) of keywords is analyzed, a fourth cluster emerges. It is provided in Figure 3.

This new cluster contains keywords such as entanglement, quantum correlation, master equation, quantum discord, phase transition, and so on, which indicates the clustering of terms on the more theoretical side. Therefore, it can be said that on a grand scale, there are three clusters of keyword associations in the academic literature of QT. As we increase our resolution, more clusters begin to become distinguishable as subfields within the overall chorus of scientific works.

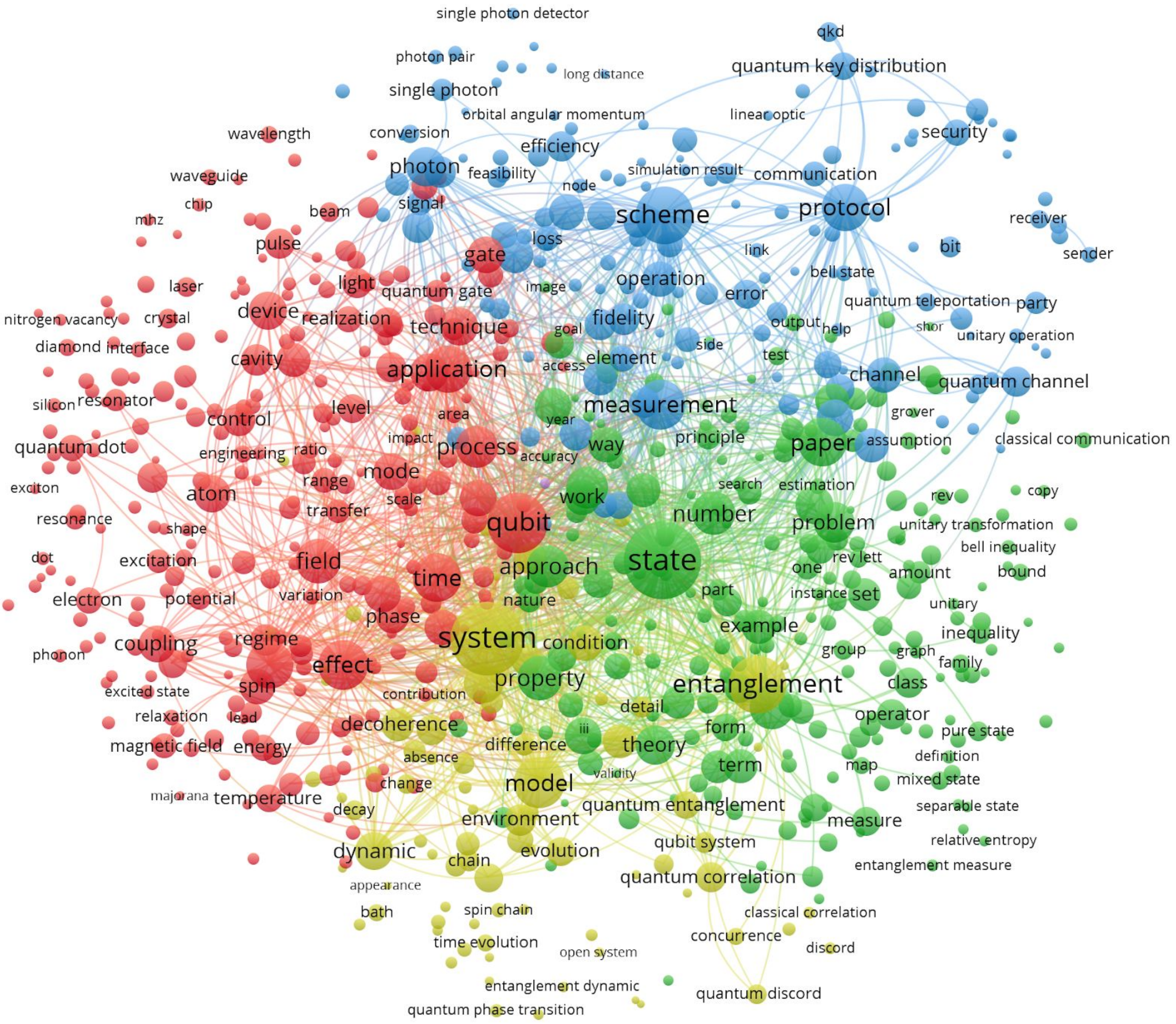

Figure 3: Clustering of terms occurring more than 200 times in the total set (100% version) (https://figshare.com/s/044516b5aeb8442b8d1f)

**(RQ2) What are the emerging and disappearing research concepts in QT research?**

This is followed by applying a burst detection analysis on the curated set to identify the emerging and disappearing research concepts in QT research. Burst detection is an algorithm that identifies sudden increases in the frequency of words by employing "a probabilistic automaton whose states correspond to increasing frequencies of individual words" (Weingart 2010). For burst analysis, we extracted relevant column pairs and harmonized the combined use of lowercase-uppercase. Only analyses over author keywords are presented here. Temporal burst detection analyses were run on the Sci2 Tools program using an algorithm for analysis of bursty and hierarchical structure in streams (Kleinberg 2007).

Table 1: Density parameter versus number of keywords for Author Keywords Column

| | | |
|---|---|---|
| *Density parameter (Gamma)* | 2.0 | 1.2 |
| *Number of keywords* | 69 | 11 |

Lowering the density parameter means only stronger bursts can be obtained. In Figure 4, outcomes for the lower density parameter can be seen. The initial one with the longest streak is quantum computation. Keywords such as quantum computer and quantum computing can also be found in the same time period. Entanglement, quantum entanglement, quantum theory are expected dominant words. The 11-year streak of teleportation as a hot topic and increasing importance of quantum discord and quantum correlations are also indicative of certain areas of interest emerging and fading away or becoming established, hence losing their high popularity among quantum researchers.

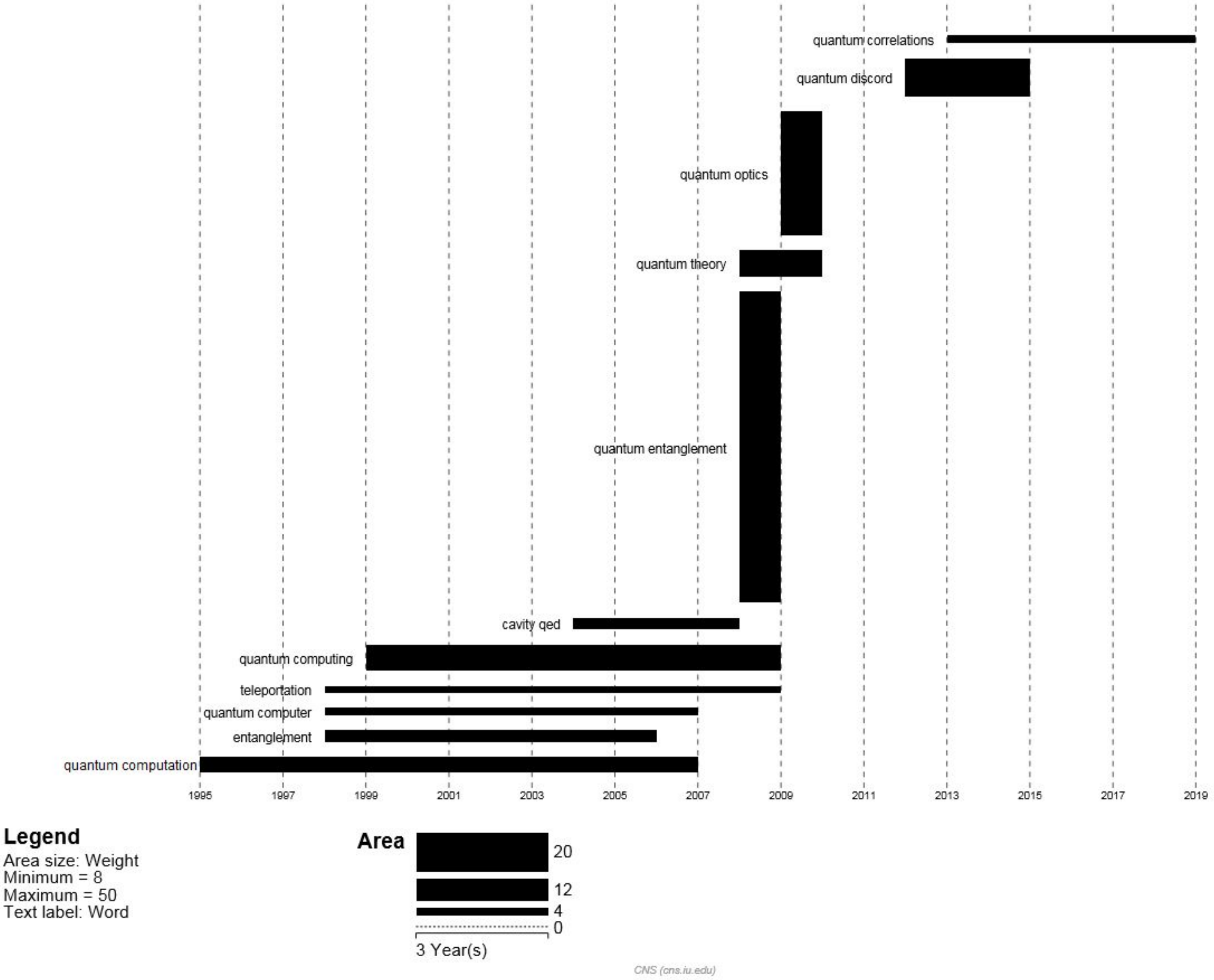

Figure 4: Results of burst analysis with density parameter 1.2

(https://figshare.com/s/9b1e7264e37d3afeddd9)

For the extended version of these topics, a higher density parameter is used, and 69 keywords obtained. Continuing streaks are provided in Table 2 with their relative weights, and the full list is in the Appendix (Table A1). This extended version gives us more information on specific topics. For example, the recent popularization of ads-cft correspondence, open quantum systems, quantum networks, quantum annealing, quantum sensing, and so on indicates a shift in literature from a more quantum computation and quantum information-theoretic subjects oriented academic literature to more diverse topics that are built upon the previous ones.

Table 2: Results of burst analysis with density parameter 2.0

| *Word* | *Weight* | *Start* | *End* |
|---|---|---|---|
| quantum coherence | 14.38 | 2016 | - |
| quantum simulation | 13.53 | 2016 | - |
| quantum image processing | 10.55 | 2015 | - |
| quantum networks | 9.99 | 2016 | - |
| quantum annealing | 8.60 | 2016 | - |
| quantum steering | 8.18 | 2016 | - |
| gauge-gravity correspondence | 7.76 | 2016 | - |
| quantum sensing | 7.71 | 2016 | - |
| ads-cft correspondence | 9.36 | 2015 | - |
| quantum metrology | 9.19 | 2015 | - |
| open quantum systems | 8.17 | 2015 | - |

At this point to double check these outcomes we ran an additional frequency analysis on some of the keywords to see whether the burst analysis results are consistent with them. Initially we ran quantum coherence, quantum simulation, and quantum annealing, which have ongoing bursts as of 2019. This is followed by ‘quantum teleportation’, which is not bursting. Finally, we check for ‘quantum computing’, which had a major burst in 2009 according to the burst analysis algorithm (check full list in Appendix).

It can be seen from figure 5 that the burst algorithm is working properly in identifying the emergence and fading away of topics. Though it is important to note the distinction that burst analysis does not provide us with the answer of whether a concept has dropped out of use or just became mainstream. In the case of ‘quantum teleportation’ it appears that the use of this term has stabilized around 0.7% frequency since the year 2000. In contrast, the term ‘quantum annealing’ has reached that level of use only in 2019. Hence, although the results of burst analysis can even be used on face value for the bursting terms, to investigate what happens to the terms after their burst ends require further analysis.

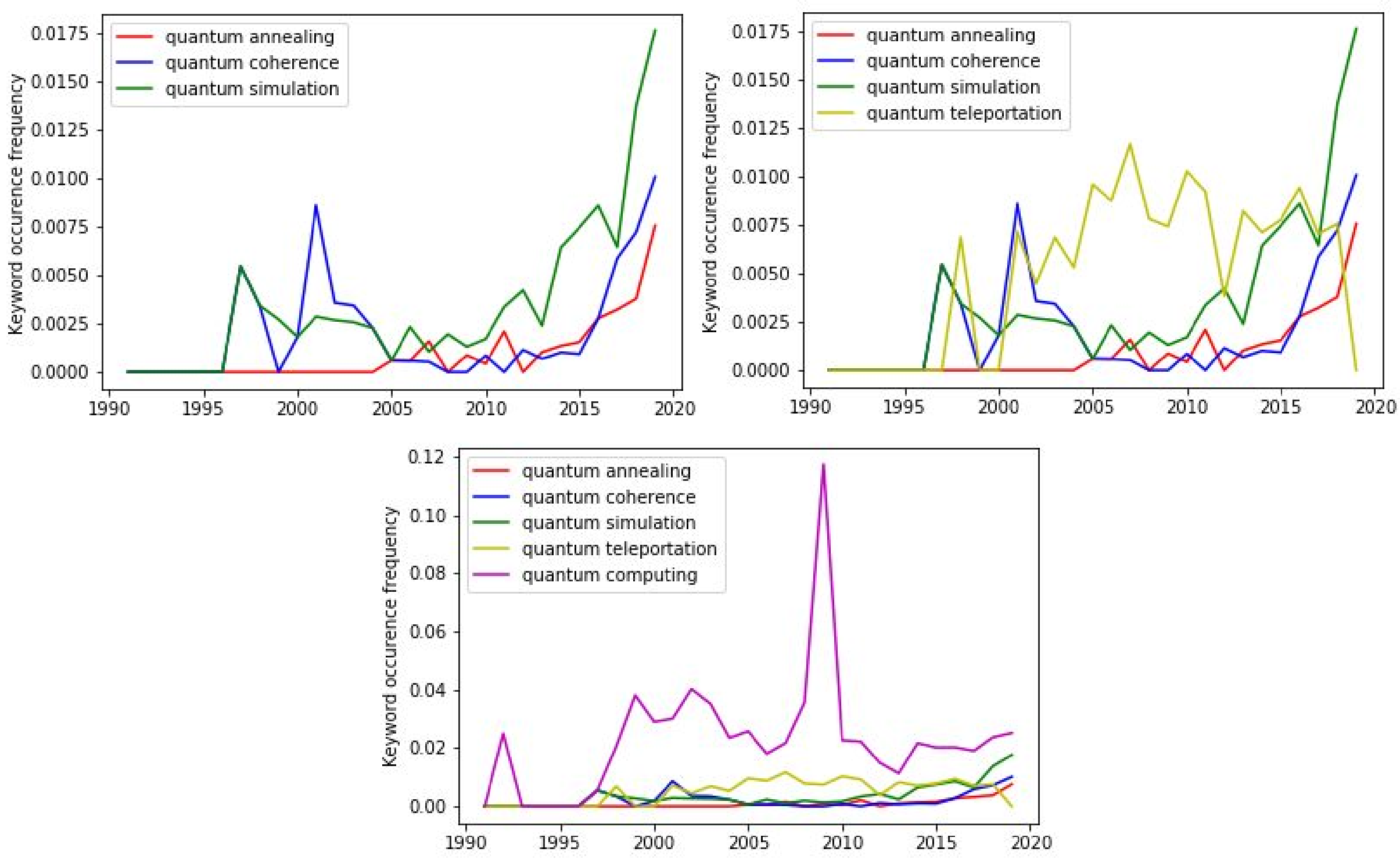

Figure 5: Keyword occurrence frequencies with respect to years in the keywords section of articles in the given year (a) quantum annealing, quantum coherence, quantum simulation (b) +quantum teleportation (c) +quantum computing (https://figshare.com/s/8008da7e377bca56afb5)

**(RQ3) Has QT research become more collaborative over the years?**

After establishing the emerging and fading away of research concepts in QT research through burst analysis we continued to answer the question whether the field was becoming more collaborative or not. This question was posed especially since Physics itself as a field has been becoming more collaborative. To extract the authorship information from WoS based data we counted the number of special characters “;” in each authors column and added one. After feeding the columns of year and number of authors and group by with respect to years we had the dataset containing only the number of authors for each article and the year it was published. To get a clearer picture initially we ran the whole set, then limited to articles with less than 50 authors, and finally to articles with less than 10 authors.

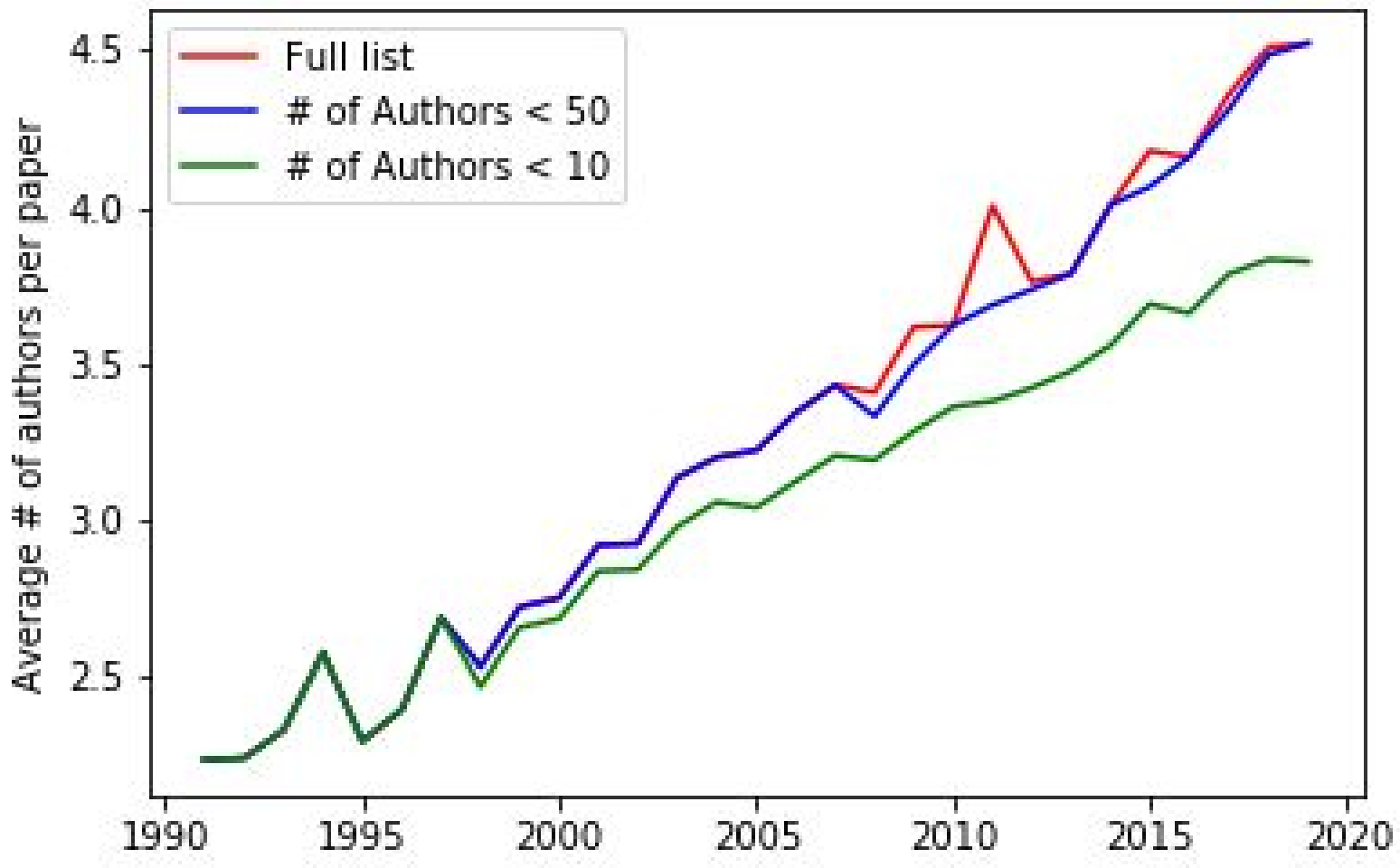

Figure 6: Average number of authors per article with respect to year. Red line represents the full list of articles, blue line for articles with less than 50 authors, and green line less than 10 (https://figshare.com/s/168b3617939900c59851)

The results presented in figure 6 shows that in all cases the average number of authors per article increased steadily almost each year, showing that the field has been becoming more collaborative. Mode of the entire set was two, which indicates that most of the articles in QT literature have two authors. Though starting from 2013 the mode of each year raises to three, giving that most of the articles published in the field since 2013 have three authors.

Until this point, we have presented our findings of the curated set and discussed over the entirety of the academic literature in quantum technologies. Our analyses show that although it is a field in topics with interdisciplinary tendencies, it is deeply rooted in physics as the academic domain of publication, and quantum computation as the early basis of academic literature. The main body of work can be split into three distinct significant clusters, which focus on (1) physical systems, (2) quantum computation together with quantum information theory, and (3) quantum cryptography and communication-related aspects. However, more and more topics in different subfields are emerging lately. Finally, publications in the field have been steadily becoming more collaborative for the last three decades and since early 2010s most articles have three authors where the average is as high as 4.5 authors.

**(RQ4) What is the core corpus of the QT literature?**

At this point we focus our attention to identify the core corpus of this field, the body of scholarly work that has been referenced by the literature most. In order to obtain not only the most cited articles but articles that have been cited most by impactful articles we acquired a second set of data. The ‘highly cited articles’ provided by Web of Science in accord with our search query; there were 808 publications. Out of these, there were 34 with over 30 citations in this set of highly cited articles. At this point, we manually obtained their respective number of citations in the curated set of 42,530 articles. The full list is given in Appendix, but here we wanted to present the top five articles with respect to citations in the highly cited articles set and curated full set.

Table 3: Citation numbers of top five core corpus in the curated set

| *cited reference* | *citations in highly cited set (808)* | *citations in curated set (42,530)* |
|---|---|---|
| nielsen m. a., 2000, quantum computation | 113 | 6929 |
| bennett ch, 1993, phys rev lett, v70, p1895, doi 10.1103/physrevlett.70.1895 | 42 | 4486 |
| ekert ak, 1991, phys rev lett, v67, p661, doi 10.1103/physrevlett.67.661 | 42 | 3137 |
| gisin n, 2002, rev mod phys, v74, p145, doi 10.1103/revmodphys.74.145 | 39 | 2262 |
| einstein a, 1935, phys rev, v47, p0777, doi 10.1103/physrev.47.777 | 46 | 2207 |

The most cited result in both cases is the essential textbook in quantum computation and quantum information fields. For the curated set, this is followed by two cornerstone articles and one highly celebrated review on quantum cryptography. The final one is the EPR paradox paper, which led to the discovery of quantum entanglement hence caused the second quantum revolution.

Table 4: Citation numbers of top five core corpus in the highly cited set

| *cited reference* | *citations in highly cited set (808)* | *citations in curated set (42,530)* |
|---|---|---|
| nielsen m. a., 2000, quantum computation | 113 | 6929 |
| kimble hj, 2008, nature, v453, p1023, doi 10.1038/nature07127 | 69 | 960 |
| wallraff a, 2004, nature, v431, p162, doi 10.1038/nature02851 | 57 | 1182 |
| nayak c, 2008, rev mod phys, v80, p1083, doi 10.1103/revmodphys.80.1083 | 56 | 699 |
| horodecki r, 2009, rev mod phys, v81, p865, doi 10.1103/revmodphys.81.865 | 49 | 1837 |

For the top five core corpus in the highly cited set, all but the most cited work are different. Kimble's article coined the term "Quantum Internet." Wallraff's article opened the doorway to building superconducting qubits of today. Nayak's article is an essential first step toward topological quantum computing, which, if realized, is expected to become the dominant method of producing qubits. Finally, Horodecki's article is an encompassing review article that incorporates contemporary approaches to quantum information theory and wide expansion of the quantum entanglement concept into differentiated aspects like quantum discord, bound entanglement, and entropic inequalities.

Comparing tables 3 and 4, the general idea emerges as, although the wider audience of academic literature is more oriented toward quantum cryptography, the highly cited works are more closely related to quantum communication, computation, and information theory. This can also be linked to the clusters identified in figure 2. Works with high citation numbers in the curated set are more likely to belong in the blue cluster (quantum cryptography and communication) while high citation numbers in the highly cited set are more closely related to the green cluster (computation and information theory).

Additionally, we have a visualization of the co-citation analysis over the highly cited articles set. Links between two articles indicate that they have been cited together, bolder the link means higher the number of co-citing articles. Coloring here is independent of figures 2 and 3.

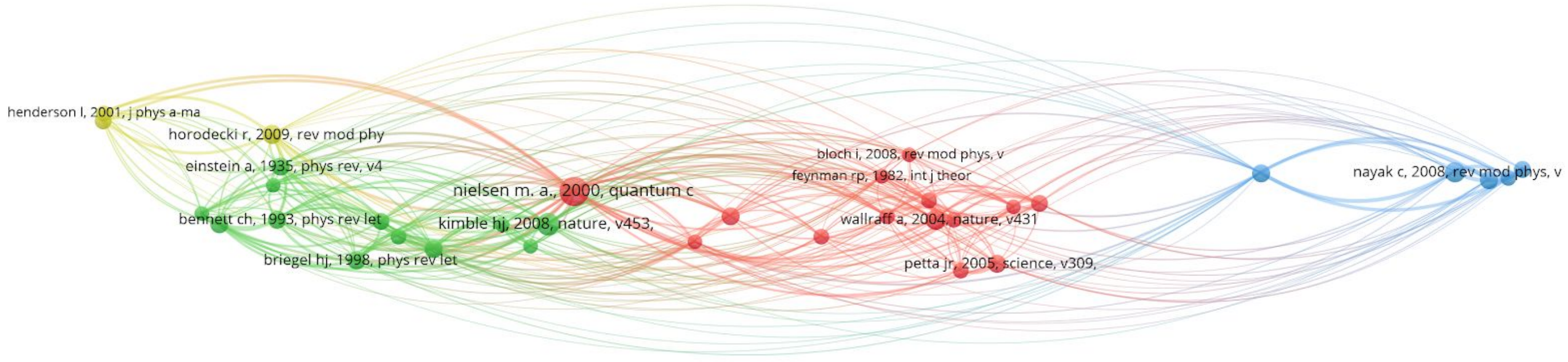

Figure 7: Co-citation analysis of the core corpus devised from the highly cited set (https://figshare.com/s/aee53c3d9aed60973c82)

The red and blue clusters correspond to articles related to quantum computing, blue ones are more related to topological quantum computing. Green cluster has articles of Bennett, Briegel, and Kimble, which are important names in quantum cryptography and communication. The yellow cluster has quantum correlations and measures, which are widely utilized for cryptography and communication purposes. This shows an alternative method of clustering the literature, instead of keywords co-citation analysis can also be performed. The main downside of applying this method to a large dataset is that it lacks the apparent explanatory power of keyword analysis. Additional explanation of each cluster needs to be attached to the visualization since it only provides the authors, year and journal.

**(RQ5) Who are the important players in QT research?**

Finally, to identify the important players in QT literature we ran co-authorship analysis on the curated set on two levels; country and institution. Figure 8 shows the results of the analysis on country level with countries that have at least 10 articles in the set, and figure 9 represents the result of the same analysis but for organizations with at least 150 articles.

Figure 8 illustrates the clustering of countries in their collaboration patterns. The largest nodes are China, the US, Germany, and England. It can be seen that each of these countries form their own clusters in terms of color but positioning of the nodes indicate that these clusters are not actually clear cut as the ones in figure 2. To get a better understanding we need to look at collaboration patterns on the institutional level.

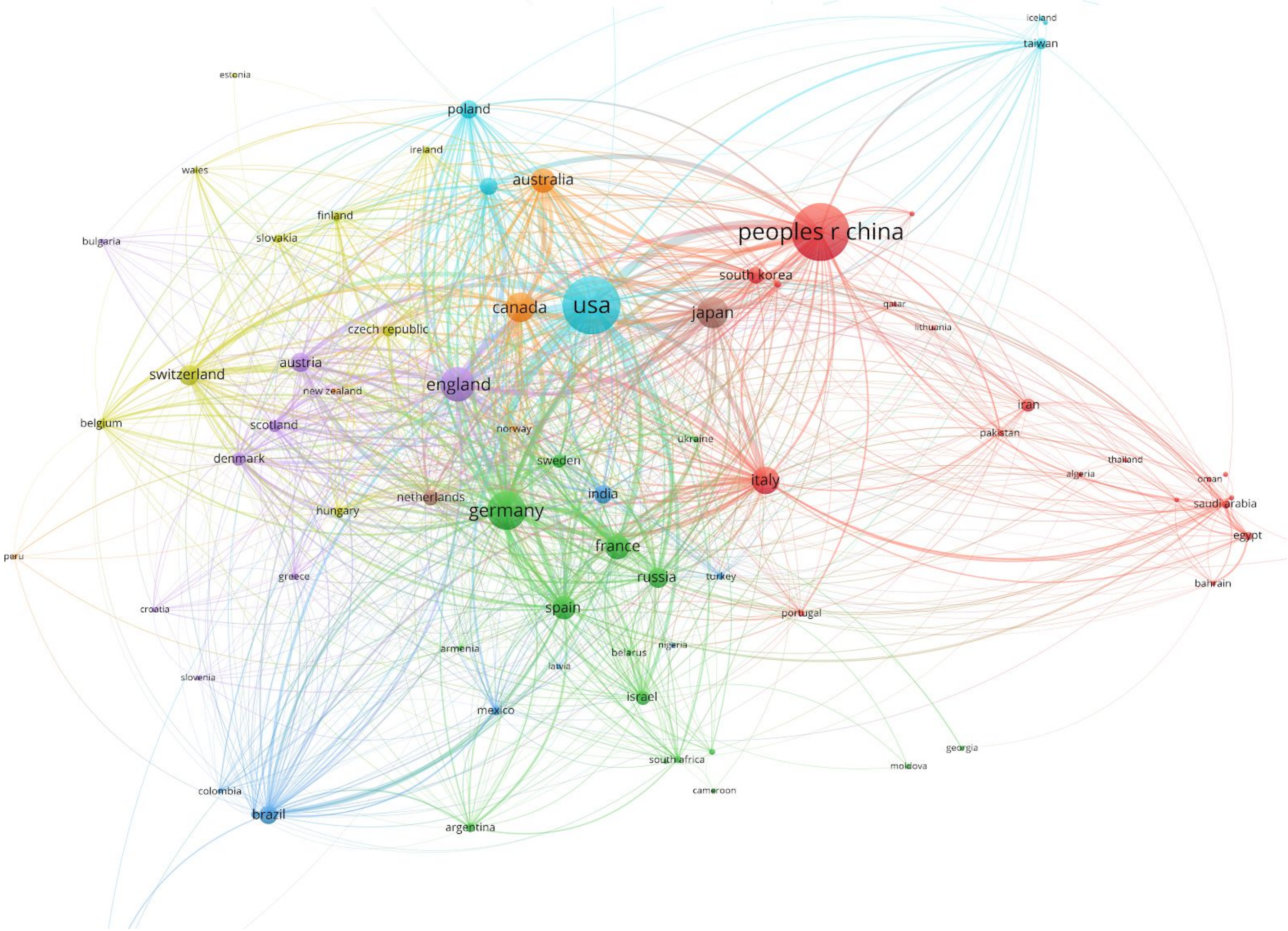

Figure 8: Co-authorship analysis of countries with more than 10 published articles (https://figshare.com/s/26bb13ebe658a6c4cc86)

On figure 9 it can be seen that collaboration patterns are more clear with respect to figure 8. This figure provides several insights. First, it appears that the internal collaboration patterns are much stronger at Chinese and Japanese institutions against their European and American counterparts. One apparent reason for this in China is that authors having multiple affiliations account as collaboration; hence, a single author article of a researcher with three affiliations would create a three node network. This effect alone would not be able to explain all separation but it is expected to be one of the reasons that the Chinese institutions are distantly separated from almost the rest of the world.

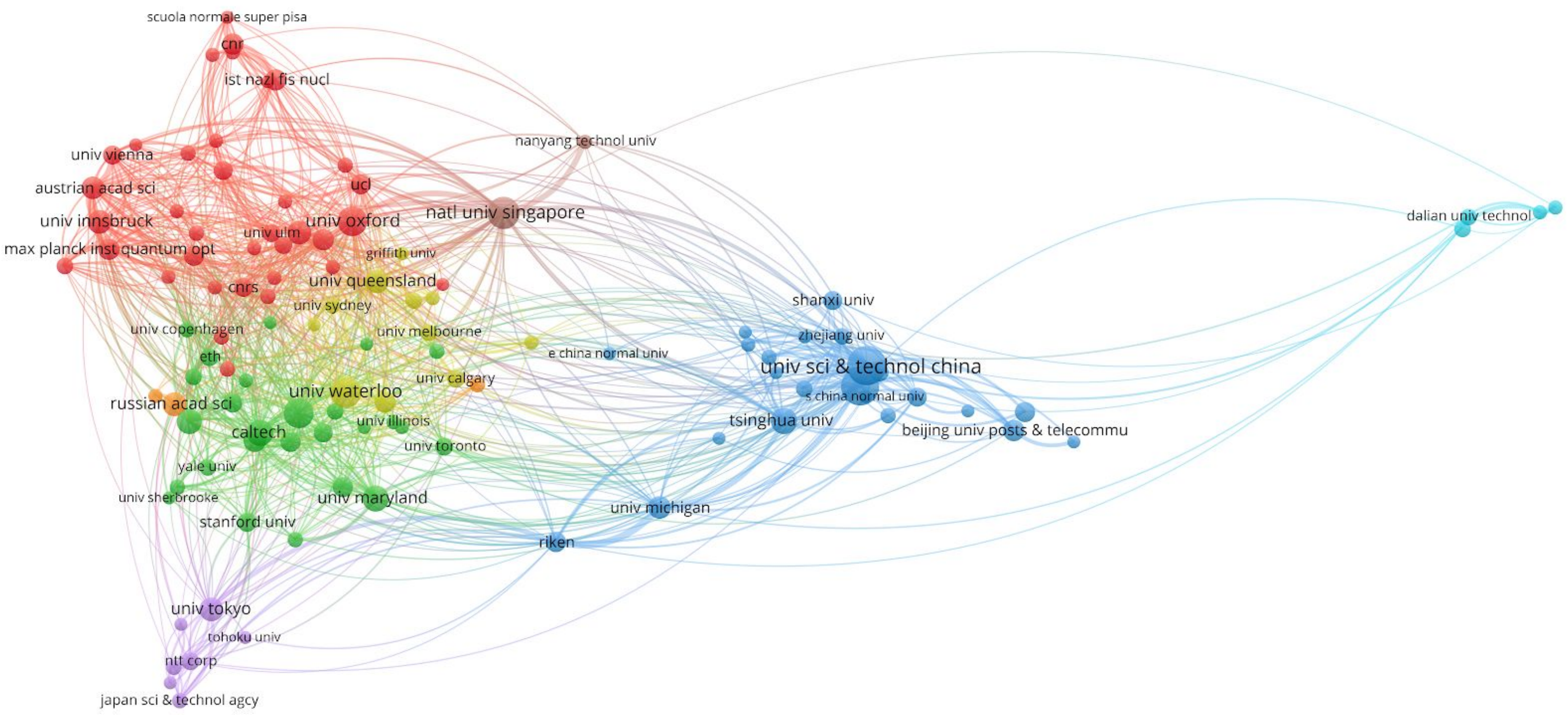

Figure 9: Co-authorship analysis of organizations with more than 150 published articles (https://figshare.com/s/557f87157ed5f2aae818)

The figure also shows that institutions in Europe collaborate with each other more often than they collaborate with institutions in the US, Japan, and China. There are several exceptions such as The University of Copenhagen and ETH Zurich, that share strong ties with American universities. Furthermore, it can be seen that Singapore and Russia form their own clusters which is expected since both countries invest in quantum technologies on national levels.

## Discussion and Conclusion

In this study, we have utilized bibliometric methods and visualization tools to investigate the literature of quantum technologies as an emerging field. We established that QT is rooted in physics, and its popularity arose through quantum computation between the late 1990s till around 2010, and following that period, the field had newer emerging topics. The main academic literature in QT is divided into three clusters; (i) quantum communication/cryptography, (ii) quantum computation/information theory, and (iii) physical realizations of quantum systems. Secondly, we have identified previous and current hot topics in the literature using burst analysis. Thirdly, we have shown that the field has been becoming more collaborative for the last three decades. Fourthly, we have formulated a core corpus of 34 articles using the co-citation analysis over 808 highly cited articles in the field. Finally, we have shown how the collaboration patterns are on country and institutional levels.

Identification of clusters in the academic literature using keyword analysis revealed two key insights. First, physical realizations of quantum systems form a cluster of their own. For example, the term quantum dot appears closely connected to the terms nitrogen vacancy, diamond, and nuclear spin, while being far away from any identifiers of the specific theoretical fields. This indicates that the separation of literature into the scholarly works of experimental and theoretical studies is almost as strong as the separation of literature into subfields of quantum computation and quantum communication. Secondly, the terms technology and quantum technology appear to be positioned at the boundary between quantum communication/cryptography and physical realizations clusters in figure 2. Although in the public sphere, the discussion on quantum technologies is mostly related to quantum computation, in the academic literature, it is not the case.

Figure 4, Table 2, and Table A1 contain the results of burst analyses, which show the emergence and fading away of concepts in the literature. It is clear that from the early days on quantum computation and concepts built around it were the central hot topics, and only after 2007, we started to see a diversification of emerging keywords. When consulted, different experts had different explanations; for example, two of the experts from quantum communication and cryptography background noted that around that time (2006-2007), there was an important development in quantum dot technology and that might have caused an experimental surge of different topics. Another expert argued that it is about funding. Both are plausible explanations; the mid-2000s were an active period for quantum technologies. In 2007, Singapore founded the Centre for Quantum Technologies (CQT), Natural Sciences and Engineering Research Council of Canada (NSERC) started funding quantum research more heavily (Sussman et al. 2019), Lockheed Martin patented a quantum radar (European patent number EP1750145), and so on. However, it is important to note that the rise in the number of articles remained steadily increasing since the early 2000s, hence a burst in the number of emerging concepts is not equivalent to a spike in the increase of the number of articles.

Following this we have constructed a core corpus of the academic literature in quantum technologies as a novel contribution. This core corpus corresponds to the fundamental and momentous articles, studies that opened up new fields or helped existing ones to further develop greatly. Two key points can be highlighted about our findings regarding the analysis of this core corpus. First, citations of these core corpus articles are different for the highly cited set and general set. This indicates that other than certain key papers which were extensively used by everyone, most papers were cited mainly by their own communities. The co-citation analysis, shown in figure 7, also shines a light to the correspondence between different subfields of quantum technologies and their respective cornerstone papers. An alternative clustering can be done with respect to using citation networks developing around these small clusters. We should also note that alternative sets of core corpora can be constructed with stricter or looser restrictions, but most of the studies included in our corpus must also appear in those versions as well.

Finally, we have shown that the field has been becoming more collaborative as the average number of authors increases steadily and the mode of number of authors has risen to three since 2013. Linking this to the collaboration patterns on country and institutional levels reveals certain insights. First, international collaboration is an important part of the academic literature in QT. Figure 8 shows this clearly as there are no distant clusters, countries that invest in this field have strong internationally collaborative patterns. However, figure 9 shows that institutions in China and Japan are more oriented towards national collaboration. A similar argument can also be presented for German and Italian institutions as well, but to a much lesser extent. Though in general, institutions from Europe and the US are in closer collaboration than their Asian counterparts.

To sum up, we accomplished four main tasks. First, we clustered the literature with respect to topics in order to enable researchers working on a topic to easily locate which other topics they might be interested in. Secondly, we identified previous and current trends in the literature to allow a wider view for decision-makers (principal investigators, funding agencies, etc.) of how the literature has been evolving since its inception. Third, we constructed a core corpus to see which articles are actually groundbreaking or field making within the context of academic literature in quantum technologies. Finally, we have analysed the collaboration dynamics of the field in number of authors, on country and institutional levels.

The study at hand can be extended in three particular aspects. First, each subfield defined by policy papers such as quantum computation, quantum sensing/imaging, quantum simulation, and quantum cryptography/communication can be studied by themselves through careful selection of query terms and curation of the obtained datasets. An example of such a study is already present for quantum cryptography research in China (Olijnyk 2018). Due to the interdisciplinary and highly entangled nature of the field, this would require expert review for each subfield. Second, the relationship between major institutions within a country and from different countries can be investigated to analyze collaboration patterns. These can be cross-checked with both investment into, and commercialization outcomes of the institutions, and policy suggestions can be developed on whether current collaboration efforts are producing the desired results or not. Thirdly, data at hand can be compared with patent analysis of quantum technologies (Winiarczyk et al. 2013; Travagnin 2019) to obtain a wider picture of the field.

The second quantum revolution is a full-blown scientific and technological phenomenon, and more studies on the emergence and development of this field are required both in academic and commercial terms for the formulation of evidence-based policy suggestions. Such efforts are now mostly reserved for non-academic journals and white papers of interested institutions, but we believe that academic studies using bibliometric tools can provide valuable insight into this ongoing technological revolution.

**Acknowledgements**

We thank our esteemed colleagues İlke Ercan, Ceyhun Bulutay, Özgür Müstecaplıoğlu, Zafer Gedik and Serkan Ateş for their contributions through the expert opinion process. Also we would like to thank Jaroslaw Miszczak and Jacob Biamonte for insightful discussions.

### Author Contributions

Z.C.S. collected and curated data, conceptualization and visualization. A.U.A. provided methodology, supervision, review and editing.

**Competing Interests:** The authors declare no competing interests.

**Funding:** The author(s) received no financial support for the research, authorship, and/or publication of this article.

**Data Availability:** Data used here is publicly available bibliographic materials collected from the Web of Science. It can be accessed via using the search queries provided and limiting the dates accordingly.

**Appendix**

Initial search query;
TS = ("quantum simulation" OR "quantum imaging" OR "quantum sensing" OR "quantum sensor" OR "quantum computation" OR "quantum computing" OR "quantum computer" OR "quantum coding" OR "quantum programming" OR "quantum error correction" OR "quantum error correcting" OR "quantum circuits" OR "quantum algorithm" OR "quantum algorithms" OR "quantum network" OR "quantum networks" OR "quantum channel" OR "quantum channels" OR "quantum cryptology" OR "quantum cryptography" OR "quantum key" OR "quantum teleportation" OR "quantum information" OR "quantum technology" OR "quantum technologies")
OR
TI = ("quantum simulation" OR "quantum imaging" OR "quantum sensing" OR "quantum sensor" OR "quantum computation" OR "quantum computing" OR "quantum computer" OR "quantum coding" OR "quantum programming" OR "quantum error correction" OR "quantum error correcting" OR "quantum circuits" OR "quantum algorithm" OR "quantum algorithms" OR "quantum network" OR "quantum networks" OR "quantum channel" OR "quantum channels" OR "quantum cryptology" OR "quantum cryptography" OR "quantum key" OR "quantum teleportation" OR "quantum information" OR "quantum technology" OR "quantum technologies")

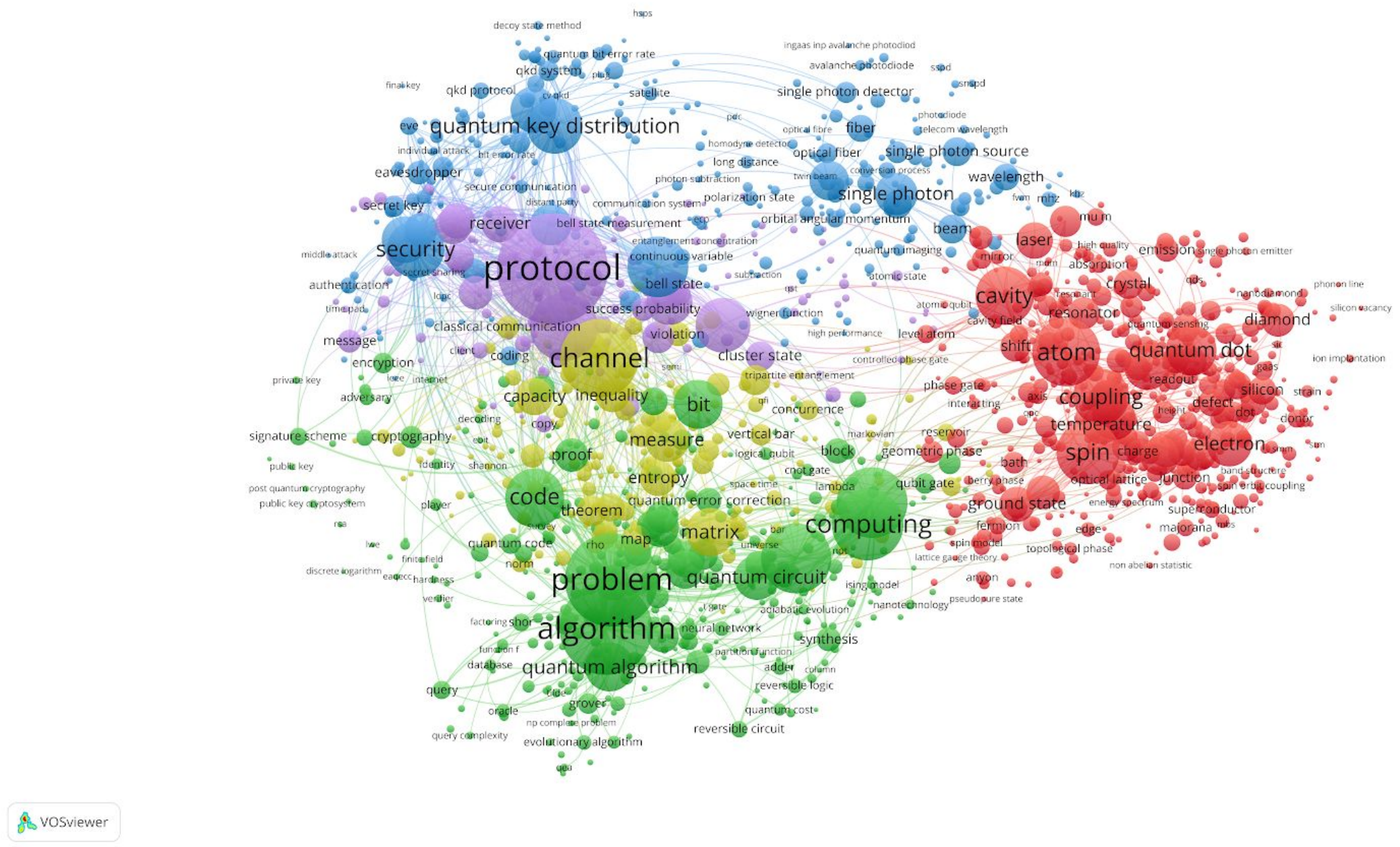

Figure 10: Clustering of terms with the dataset generated by the early search on March 2019 with the initial search query ( https://figshare.com/s/adde250d9c0a7dc446ca)

Final search query;
TS = ("quantum simulation" OR "quantum imaging" OR "quantum sensing" OR "quantum sensor" OR "quantum computation" OR "quantum computing" OR "quantum computer" OR "quantum coding" OR "quantum programming" OR "quantum error correction" OR "quantum error correcting" OR "quantum circuits" OR "quantum algorithm" OR "quantum algorithms" OR "quantum network" OR "quantum networks" OR "quantum channel" OR "quantum channels" OR "quantum cryptology" OR "quantum cryptography" OR "quantum key" OR "quantum teleportation" OR "quantum information" OR "quantum technology" OR "quantum technologies" OR “quantum gates" OR “quantum register" OR “quantum contextuality" OR “quantum decoherence" OR “quantum communication" OR “quantum memory" OR “quantum memories" OR “quantum repeaters" OR “quantum state transfer" OR “quantum zeno dynamics" OR “qubit" OR “qutrit" OR “qudit" OR “quantum correlations" OR “quantum entanglement" OR “quantum discord" OR “quantum noise engineering" OR “quantum state engineering" OR “quantum protocols" OR “quantum annealing” OR “quantum logic gate” OR “quantum internet” OR “quantum repeater” OR “quantum memory” OR “quantum photonics” OR “quantum photonic” OR “quantum biology” OR “quantum machine learning”)
OR
TI=("quantum simulation" OR "quantum imaging" OR "quantum sensing" OR "quantum sensor" OR "quantum computation" OR "quantum computing" OR "quantum computer" OR "quantum coding" OR "quantum programming" OR "quantum error correction" OR "quantum error correcting" OR "quantum circuits" OR "quantum algorithm" OR "quantum algorithms" OR "quantum network" OR "quantum networks" OR "quantum channel" OR "quantum channels" OR "quantum cryptology" OR "quantum cryptography" OR "quantum key" OR "quantum teleportation" OR "quantum information" OR "quantum technology" OR "quantum technologies" OR “quantum gates" OR “quantum register" OR “quantum contextuality" OR “quantum decoherence" OR “quantum communication" OR “quantum memory" OR “quantum memories" OR “quantum repeaters" OR “quantum state transfer" OR “quantum zeno dynamics" OR “qubit" OR “qutrit" OR “qudit" OR “quantum correlations" OR “quantum entanglement" OR “quantum discord" OR “quantum noise engineering" OR “quantum state engineering" OR “quantum protocols" OR “quantum annealing” OR “quantum logic gate” OR “quantum internet” OR “quantum repeater” OR “quantum memory” OR “quantum photonics” OR “quantum photonic” OR “quantum biology” OR “quantum machine learning”)

Table A1: Results of burst analysis with density parameter 2.0

| Word | Weight | Start | End |
|---|---:|---:|---:|
| ads-cft correspondence | 9.36 | 2015 | - |
| open quantum systems | 8.17 | 2015 | - |
| quantum networks | 9.99 | 2016 | - |
| quantum annealing | 8.6 | 2016 | - |
| quantum sensing | 7.71 | 2016 | - |
| quantum image processing | 10.55 | 2015 | - |
| gauge-gravity correspondence | 7.76 | 2016 | - |
| quantum metrology | 9.19 | 2015 | - |
| quantum simulation | 13.53 | 2016 | - |
| quantum steering | 8.18 | 2016 | - |
| quantum coherence | 14.38 | 2016 | - |
| weak measurement | 12.87 | 2014 | 2016 |
| quantum correlation | 13.03 | 2013 | 2016 |
| entanglement concentration | 9.34 | 2012 | 2015 |
| quantum discord | 49.81 | 2012 | 2015 |
| quantum correlations | 9.66 | 2013 | 2015 |
| cluster state | 8.33 | 2010 | 2013 |
| thermal entanglement | 10.53 | 2006 | 2013 |
| quantum state sharing | 11.48 | 2010 | 2013 |
| quantum secret sharing | 16.3 | 2010 | 2013 |
| iii-v semiconductors | 12.53 | 2008 | 2010 |
| elemental semiconductors | 8.48 | 2008 | 2010 |
| gallium arsenide | 10.67 | 2008 | 2010 |
| quantum theory | 26.96 | 2008 | 2010 |
| semiconductor quantum dots | 10.63 | 2008 | 2010 |
| ising model | 9.93 | 2009 | 2009 |
| boson systems | 9.81 | 2008 | 2009 |
| fermion systems | 7.75 | 2009 | 2009 |
| probability | 19.46 | 2008 | 2009 |
| eigenvalues and eigenfunctions | 13.08 | 2009 | 2009 |
| entropy | 11.61 | 2009 | 2009 |
| encoding | 9.25 | 2008 | 2009 |
| matrix algebra | 13.61 | 2008 | 2009 |
| quantum electrodynamics | 10.57 | 2008 | 2009 |
| quantum entanglement | 161.8 | 2009 | 2009 |
| hilbert spaces | 16.72 | 2008 | 2009 |
| quantum computing | 103.75 | 2009 | 2009 |
| information theory | 25.35 | 2008 | 2009 |
| excited states | 7.73 | 2008 | 2009 |
| mathematical operators | 8.4 | 2008 | 2009 |
| teleportation | 26.72 | 2001 | 2009 |
| protocols | 18.54 | 2008 | 2009 |
| measurement theory | 10.01 | 2008 | 2009 |
| optical squeezing | 12.81 | 2009 | 2009 |
| quantum gates | 24.18 | 2009 | 2009 |
| ground states | 23.18 | 2008 | 2009 |
| bell theorem | 16.88 | 2008 | 2009 |
| atom-photon collisions | 16.39 | 2009 | 2009 |
| error correction codes | 7.87 | 2008 | 2009 |
| squids | 10.41 | 2009 | 2009 |
| quantum optics | 68.33 | 2009 | 2009 |
| radiation pressure | 8.45 | 2009 | 2009 |
| two-photon processes | 10.38 | 2008 | 2009 |
| qubit | 12.44 | 2001 | 2008 |
| cavity qed | 21.17 | 2006 | 2008 |
| quantum logic | 9.48 | 1998 | 2008 |
| probabilistic teleportation | 8.94 | 2002 | 2008 |
| unitary transformation | 8.21 | 2004 | 2007 |
| quantum computer | 33.38 | 1998 | 2007 |
| josephson junction | 12.7 | 2000 | 2007 |
| quantum computation | 80.44 | 1995 | 2007 |
| nmr | 10.5 | 1997 | 2007 |
| decoherence | 10.43 | 2003 | 2006 |
| quantum state engineering | 8.96 | 2004 | 2006 |
| spintronics | 9.49 | 2000 | 2005 |
| quantum computers | 11.18 | 1997 | 2005 |
| entanglement | 19.92 | 2001 | 2005 |
| quantum information theory | 9.73 | 1996 | 2003 |
| quantum computing | 17.56 | 1999 | 2002 |

Table A2: Citation numbers of core corpus articles in highly cited and total sets

| cited reference | citations in 808-hc | citations in 42,530 |
|---|---|---|
| balasubramanian g, 2009, nat mater, v8, p383, doi [10.1038/nmat2420 10.1038/nmat2420] | 32 | 398 |
| bell j.s., 1964, physics, v1, p195 | 31 | 1661 |
| bennett c h, 1984, p ieee int c comp sy, p175 | 30 | 2136 |
| bennett ch, 1993, phys rev lett, v70, p1895, doi 10.1103/physrevlett.70.1895 | 42 | 4486 |
| blais a, 2004, phys rev a, v69, doi 10.1103/physreva.69.062320 | 34 | 944 |
| bloch i, 2008, rev mod phys, v80, p885, doi 10.1103/revmodphys.80.885 | 30 | 335 |
| briegel hj, 1998, phys rev lett, v81, p5932, doi 10.1103/physrevlett.81.5932 | 43 | 1040 |
| clarke j, 2008, nature, v453, p1031, doi 10.1038/nature07128 | 33 | 582 |
| dicarlo l, 2009, nature, v460, p240, doi 10.1038/nature08121 | 33 | 387 |
| duan lm, 2001, nature, v414, p413, doi 10.1038/35106500 | 42 | 1147 |
| einstein a, 1935, phys rev, v47, p0777, doi 10.1103/physrev.47.777 | 46 | 2207 |
| ekert ak, 1991, phys rev lett, v67, p661, doi 10.1103/physrevlett.67.661 | 42 | 3137 |
| feynman rp, 1982, int j theor phys, v21, p467, doi 10.1007/bf02650179 | 30 | 1019 |
| fu l, 2008, phys rev lett, v100, doi 10.1103/physrevlett.100.096407 | 44 | 337 |
| gisin n, 2002, rev mod phys, v74, p145, doi 10.1103/revmodphys.74.145 | 39 | 2262 |
| henderson l, 2001, j phys a-math gen, v34, p6899, doi 10.1088/0305-4470/34/35/315 | 34 | 976 |
| hong ck, 1987, phys rev lett, v59, p2044, doi 10.1103/physrevlett.59.2044 | 36 | 690 |
| horodecki r, 2009, rev mod phys, v81, p865, doi 10.1103/revmodphys.81.865 | 49 | 1837 |
| kimble hj, 2008, nature, v453, p1023, doi 10.1038/nature07127 | 69 | 960 |
| kitaev ay, 2003, ann phys-new york, v303, p2, doi 10.1016/s0003-4916(02)00018-0 | 45 | 1003 |
| knill e, 2001, nature, v409, p46, doi 10.1038/35051009 | 30 | 1570 |
| koch j, 2007, phys rev a, v76, doi 10.1103/physreva.76.042319 | 40 | 607 |
| kok p, 2007, rev mod phys, v79, p135, doi 10.1103/revmodphys.79.135 | 31 | 635 |
| ladd td, 2010, nature, v464, p45, doi 10.1038/nature08812 | 41 | 661 |
| loss d, 1998, phys rev a, v57, p120, doi 10.1103/physreva.57.120 | 32 | 1725 |
| mourik v, 2012, science, v336, p1003, doi 10.1126/science.1222360 | 33 | 360 |
| nayak c, 2008, rev mod phys, v80, p1083, doi 10.1103/revmodphys.80.1083 | 56 | 699 |
| nielsen m. a., 2000, quantum computation | 113 | 6929 |
| ollivier h, 2002, phys rev lett, v88, doi 10.1103/physrevlett.88.017901 | 37 | 1332 |
| paik h, 2011, phys rev lett, v107, doi 10.1103/physrevlett.107.240501 | 30 | 349 |
| petta jr, 2005, science, v309, p2180, doi 10.1126/science.1116955 | 42 | 832 |
| read n, 2000, phys rev b, v61, p10267, doi 10.1103/physrevb.61.10267 | 37 | 281 |
| togan e, 2010, nature, v466, p730, doi 10.1038/nature09256 | 30 | 305 |
| wallraff a, 2004, nature, v431, p162, doi 10.1038/nature02851 | 57 | 1182 |